# Design and Modeling of $CdGa_2Te_4$ and $ZnGa_2Te_4$ Chalcogenide Compound-Based Photovoltaic Devices: A DFT Study along with SCAPS-1D Simulation

Md Hasan Shahriar Rifat*[1]; Tanvir Khan[2]; Md Arafat Hossain Shourov[1]; Md Sahat Bin Sayed[1]; Md. Saiful Islam

[1]Dept. of Materials Science and Engineering, University of Rajshahi, Rajshahi 6205, Bangladesh

[2]Dept. of Physics, University of Rajshahi, Rajshahi 6205, Bangladesh

*Corresponding author E-mail address: rifatmse@gmail.com (Md Hasan Shahriar Rifat)

## Abstract

The electronic and optical properties of $CdGa_2Te_4$ and $ZnGa_2Te_4$ were calculated by the first-principles DFT method in terms of the GGA-PBESol approximation. Two materials possess an appropriate band gap, a high absorption coefficient (~$10^4$ cm$^{-1}$) in the visible range region, a low exciton binding energy (18.85-26.81 meV), and a large Bohr radius (23-34.3Å) of moderate exciton temperature 218−311 K for PV applications. The device performance was also numerically simulated using the SCAPS-1D software for Pt/CdS/$CdGa_2Te_4$/$Cu_2O$/Ti and Pt/CdS/$ZnGa_2Te_4$/$Cu_2O$/Ti structures. The influence of the layer thickness, carrier concentration, and defect density on the solar cell characteristics was studied. The best absorber layer thickness is 1000–1800 nm, while the CdS buffer layer should be about 100 nm. A defect density of less than $1.772 \times 10^{13}$ cm$^{-3}$ is required for an efficiency larger than 20%. Simulated power conversion efficiencies were 18.46% for $CdGa_2Te_4$ and 17.35% for $ZnGa_2Te_4$ solar cells.

**Keywords:** DFT, quaternary chalcogenide compounds, SCAPS-1D, electronic properties, optical properties.

## 1. Introduction

The exclusive reliance on traditional fossil fuels as the primary energy source for achieving sustainable development is no longer viable, given the escalating global energy demand and its detrimental impact on the environment [1–5]. In response to rising environmental pollution and global energy challenges, solar energy has emerged as a popular renewable and clean alternative [6,7]. Over the past decade, perovskite solar cells (PSCs) have demonstrated significant potential for solar energy applications. The light-harvesting layer in PSCs offers a straightforward fabrication process, tunable bandgap, and extended carrier lifespan [8,9]. The power conversion efficiency (PCE) of PSCs has increased remarkably from 3.8% to 25.73% [10,11]. However, the widespread use of lead (Pb)-based compounds in solar cells (SCs) presents considerable environmental and health risks [12–14]. Therefore, there is an urgent need to develop highly efficient and less hazardous alternatives to Pb in PSCs [15]. The outstanding optoelectronic properties of halide perovskite materials have made remarkable progress possible. These properties encompass a high absorption coefficient, adjustable band gaps, long-range charge diffusion lengths, low exciton binding energies, and high charge carrier mobilities [16,17].

Recently, density functional theory (DFT) has become an essential and powerful tool for investigating the geometrical, optical, mechanical, and electrical properties of different compounds, particularly in the context of photovoltaic (PV) applications [18–20]. A notable study by Jahangirli et al. exemplifies this, as they conducted both experimental and DFT-based analyses on the electronic, optical, and vibrational properties of $CdGa_2Te_4$ using the GGA functional, underscoring its potential for PV technologies [21]. Within this broader field, a subgroup of materials known as defect chalcopyrites (DCs) has garnered significant attention due to their intriguing characteristics. These materials, structurally related to chalcopyrites, crystallize in a tetragonal structure under ambient conditions and possess the general chemical formula $AB_2X_4$, where A = Zn, Cd, or Hg; B = Ga or Al; and X = S, Se, or Te. DC semiconductors are particularly noteworthy due to their exceptional electronic, elastic, optical, and nonlinear properties, rendering them promising candidates for a wide range of technological applications, including nonlinear optics, photovoltaic devices, and photodetectors [22–24]. Among these, $CdGa_2Te_4$ and $ZnGa_2Te_4$ are distinguished as prominent members of the defect chalcopyrite family, recognized for their nearly ideal chalcopyrite-like crystal structures.

To comprehend the various characteristics of defect chalcopyrite (DC) semiconductors, several experimental and theoretical investigations have been conducted. The structural, electrical, optical, and nonlinear features of $CdGa_2X_4$ (X = S, Se) [25][26], $XAl_2Se_4$ (X = Zn, Cd, Hg) [27–30], $ZnIn_2Te_4$ [30], and $ZnGa_2Te_4$ [31] have all been examined by first principle calculations. Jiang and Lambrecht [32] have examined DC semiconductors' electronic band structures and compared them to those of their parent chalcopyrite materials. It has been theoretically computed how the structural and electrical characteristics of $CdAl_2Se_4$ vary with temperature and pressure [33]. Despite the fact that these materials have been the focus of numerous research projects, many of their basic characteristics remain unclear or inadequately assessed.

For high-performance SCs, it is important to identify the role of each functional layer, such as the absorber, hole transport layer (HTL), electron transport layer (ETL), and metal contact, in overall device performance. The absorption of light by the light-harvesting layer is particularly important among them to achieve optimum photovoltaic performance. Both the ETL and HTL are crucial, for they also control the charge carrier transport and separation that are produced by photon absorption in the absorber. The open-circuit voltage ($V_{OC}$) is due to the energy levels difference between ETL and HTL; enhanced charge mobility contributes to both fill factor (FF) and short-circuit current density ($J_{SC}$). Cadmium sulfide has been chosen as the ETL, while copper(I) oxide ($Cu_2O$) was selected as the HTL due to its suitable band gap for solar energy absorption, high carrier charge mobility, and good optical transparency. The framework presented here facilitates a systemic understanding of which components, the ETL (CdS), varying absorbers, and selected HTL ($Cu_2O$), mutually contribute to efficiency, performance, and long-term stability [34]. $Cu_2O$ was selected here in particular based on extensive literature evidence already describing it as a prime HTM material, offering high device quality in various device setups. Indeed, a high-throughput study of 72 device architectures involving 8 ETLs and 9 HTLs verified $Cu_2O$ as the most appropriate HTL material due to its compatibility, as demonstrated with a wide variety of ETL materials and efficacy [37][38].

In this paper, we report another numerical study on a Pt/CdS/$CdGa_2Te_4$/$Cu_2O$/Ti and Pt/CdS/$ZnGa_2Te_4$/$Cu_2O$/Ti solar cell with a specific Micro-Band Offset (MBO) energy alignment. In this work, we use theoretical SCAPS-1D simulation and DFT to provide both the theoretic analysis and the guide for device optimization. Maximum PCEs of 18.46% and 17.35% for the $CdGa_2Te_4$- and $ZnGa_2Te_4$-based solar cells can be obtained because the MBO energy structure creates a built-in electric field, which assists carrier transport and separation. This is primarily used to realize the perfect energy band matchings between absorber layers and ETL/HTL, high carrier transfer rate, and minimum recombination. To gain better insights into how power conversion can be controlled, a back metal electrode (BME) was introduced, and the thickness of absorber layers and their defect density and dope densities were changed. In summary, our work demonstrates that the MBO-fabricated heterostructures could serve as a promising pathway toward the next generation of thin-film solar cells with better carrier dynamics and higher photovoltaic performance.

## 2. Materials and methodology

### 2.1 First principal calculations of $CdGa_2Te_4$ and $ZnGa_2Te_4$

First-principles investigations were carried out using both the Cambridge Serial Total Energy Package (CASTEP) [35] within the framework of density functional theory (DFT) [37]. The pseudo-potential plane-wave (PP-PW) [36] approach was employed for total energy minimization to identify the most thermodynamically stable (ground state) crystal structures of $CdGa_2Te_4$ and $ZnGa_2Te_4$ compounds. Structural relaxations were conducted under the Generalized Gradient Approximation (GGA) [37], utilizing the Perdew-Burke-Ernzerhof revised for solids (PBEsol) exchange-correlation functional [38]. Convergence criteria were rigorously enforced, with an energy tolerance set at $2 \times 10^{-6}$ eV per atom and a plane-wave cutoff energy of 600 eV. The Brillouin zone was sampled using a Monkhorst-Pack grid of $8 \times 8 \times 4$ k-points. During the optimization process, constraints included a maximum atomic displacement of $5 \times 10^{-4}$ Å, a force tolerance of $5 \times 10^{-6}$ eV/Å per atom, and a maximum stress threshold of 0.02 GPa. Both $CdGa_2Te_4$ and $ZnGa_2Te_4$ were confirmed to crystallize in the tetragonal *I*-4 space group (International Space Group No. 82), characterized by lattice constants $a = b = 6.22$ Å and $c = 11.92$ Å, with all lattice angles fixed at 90°. The unit cell arrangement is illustrated in **Figure 1**. To investigate the electronic and optical properties, initial calculations were performed using the GGA-PBESol function. This multilevel approach enabled a more reliable prediction of the band gaps and optoelectronic behavior of $CdGa_2Te_4$ and $ZnGa_2Te_4$.

In this work, SCAPS-1D (Solar Cell Capacitance Simulator-1D) was used as a simulation platform due to its high efficiency and flexibility, with which the electrical characteristics of the solar cells were simulated [39]. In order to estimate the performance of the solar cells, the electron transport layer (ETL) was the light absorption layer CdS, the light absorber layer was $CdGa_2Te_4$ (or $ZnGa_2Te_4/CdGa_2Te_4$ alloy), and the hole transport layer (HTL) was $Cu_2O$. The solar cell structure used is a typical Ti/ETL/absorber/HTL/Pt structure, where Ti serves as the back contact, CdS as the ETL, $CdGa_2Te_4$ as the absorber layer, $Cu_2O$ as the HTL, and Pt as the front electrode connected to the load. For the top performances (fill factor (FF), power conversion efficiency (PCE), energy band alignment, quantum efficiency (QE), short circuit current density (Jsc), and current-voltage (J-V) characteristics), we used the SCAPS-1D simulation. These factors are of critical importance to analyze the overall efficacy, stability, and efficiency of the constructed solar cells [40]. This enables the effect of absorber material and transport layers on the device performance to be analyzed.

# 3. Result and Discussion

## 3.1 Structural properties and dynamic Stability

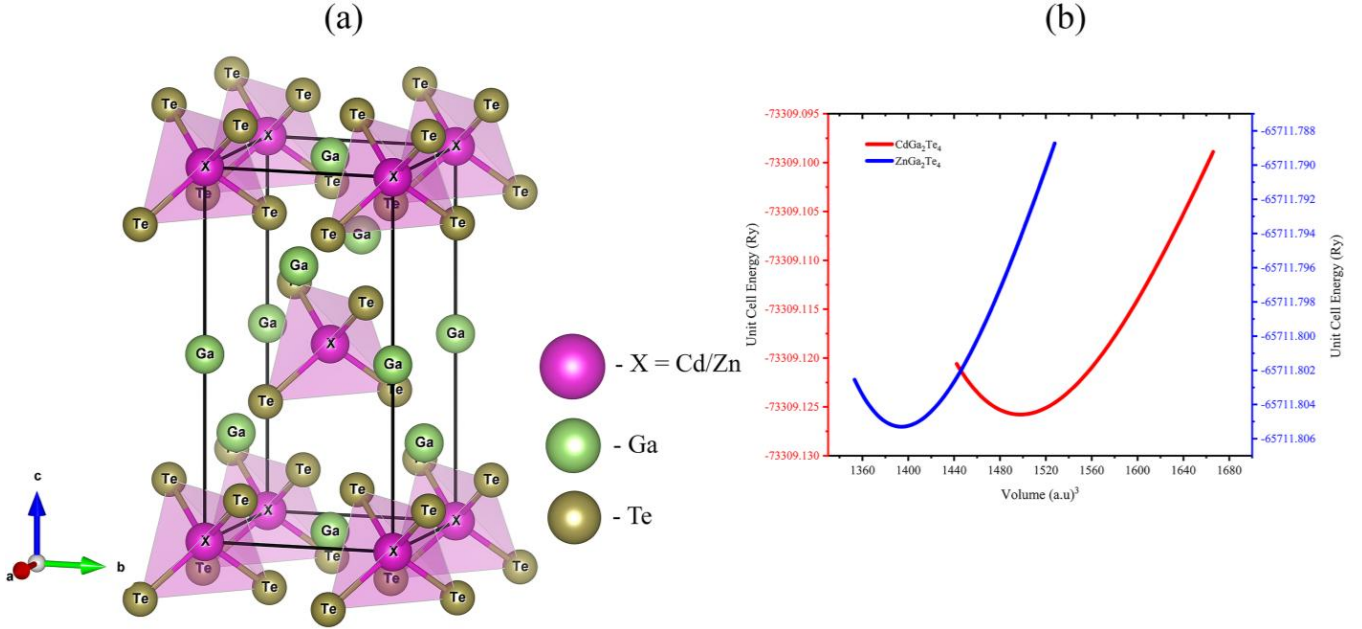

**Figure 1.** Optimized (a) crystal structure and (b) volume-optimized (E-V) plot of XGa$_2$Te$_4$ (X=Cd, Zn)

The semiconductor compounds CdGa$_2$Te$_4$ and ZnGa$_2$Te$_4$ crystallize in the tetragonal I-4 (Space Group No. 82). Figure 1a illustrates the various structural configurations of these materials that have been studied using VESTA software [41]. Ga atoms divide between 2b sites at (0.5, 0.5, 0.0) and 2d sites at (0.5, 0.0, 0.25), whereas Cd/Zn atoms inhabit the 2a Wyckoff site with coordinates (0.0, 0.0, 0.0) in these compounds. The atoms are listed in Table 1 and occupy the 8g position (0.226747, 0.755192, 0.362783). By minimizing their total energy, the equilibrium crystal structures of ZnGa$_2$Te$_4$ and CdGa$_2$Te$_4$ are found. To uncover the ground-state characteristics of XGa$_2$Te$_4$ compounds, geometry optimization is carried out at zero temperature and pressure. Through geometry optimization, the lattice parameters and cell volume are determined, as shown in Table 2. The precision and dependability of this study are further supported by this high agreement. The unit cell stable state energy, measured in electron volts (eV), was used to determine the equilibrium unit cell volume, $V_0$. This investigation involved fitting the observed energy and volume data to the Birch-Murnaghan states in **Eq. 1**. **Figure 1b** depicts the relationship between energy and volume, which allowed for the accurate determination of the stable unit cell parameters [42].

$$E(V) = E_0(V) + \left[\frac{B_0}{B'_0(B'_0 - 1)}\right] \times \left[B_0\left(1 - \frac{V_0}{V}\right) + \left(\frac{V_0}{V}\right)^{B'_0} - 1\right] \quad (1)$$

**Table 1**: Wyckoff positions of the different atomic species in XGa₂Te₄ (X=Cd, Zn) crystal structures.

| Compound | Wyckoff | Atom | x | y | z |
|---|---|---|---|---|---|
| CdGa₂Te₄ | 2a | Cd | 0 | 0 | 0 |
| | 2b | Ga | 1/2 | 1/2 | 0 |
| | 2d | Ga | 1/2 | 0 | 1/4 |
| | 8g | Te | 0.226747 | 0.755192 | 0.362783 |
| ZnGa₂Te₄ | 2a | Zn | 0 | 0 | 0 |
| | 2b | Ga | 1/2 | 1/2 | 0 |
| | 2d | Ga | 1/2 | 0 | 1/4 |
| | 8g | Te | 0.761913 | 0.265063 | 0.132453 |

**Table 2:** Lattice parameters, cell volume, formation energy, and cohesive energy of the XGa₂Te₄ (X=Cd, Zn) structure.

| Compounds | Total Energy | $a(\text{Å})$ | $c(\text{Å})$ | c/a | $\Delta E_f$ (eV/atom) | $\Delta E_c$ (eV/atom) | Ref. |
|---|---|---|---|---|---|---|---|
| CdGa₂Te₄ | -16075.038 | 6.22 | 11.92 | 1.92 | -1.44 | -3.634 | This work |
| ZnGa₂Te₄ | -17210.346 | 6.014 | 11.92 | 1.98 | -1.39 | -3.756 | This work |
| CdGa₂Te₄ | --- | 6.04 | 12.135 | 2.01 | --- | --- | (exp.) [21] |

Formation energy and cohesive energy, indicating the energy necessary to disrupt the bonds between various atoms in a crystal structure, were calculated for XGa₂Te₄ materials using the following **Eqs. 2 and 3**.

$$\Delta E_f(XGa_2Te_4) = \frac{E_{total} - (2E_{gr}(X) + 4E_{gr}(Ga) + 8E_{gr}(Te))}{14} \quad (2)$$

$$\Delta E_c(XGa_2Te_4) = \frac{E_{total} - (2E_{iso}(X) + 4E_{iso}(Ga) + 8E_{iso}(Te))}{14} \quad (3)$$

In these expressions, the total energy is the energy for the total system itself and taken as a reference level, while $E_X$, $E_{Ga}$, and $E_{Te}$ are energies of X=Cd/Zn, Ga, and Te atoms in their most stable crystal structures. From **Table 2**, the formation energies of all considered structures are negative. This implies that both structures are thermodynamically stable and synthesizable.

The physical behavior of crystalline materials is greatly influenced by phonons, which are quanta of lattice vibrations. They also have significant effects on their lattice, elastic, and electrical-thermal dynamics. While electron–phonon interaction can be linked to PDOS, the phonon spectrum is sensitive to the mass of its elements, stiffness, and crystal symmetry [43–45]. Phonon dispersion spectra (PDS) can be used to predict thermal vibration behavior, phase transitions, and dynamic stability.

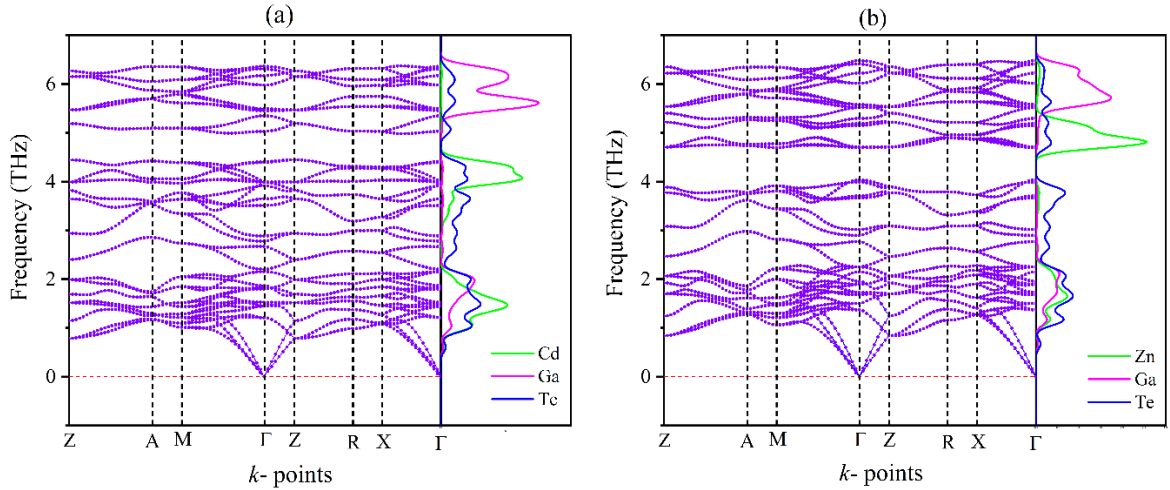

**Figure 2.** Calculated phonon spectra (phonon dispersion on the left and phonon PDOS on the right) of each curve for (a) $CdGa_2Te_4$ and (b) $ZnGa_2Te_4$.

In this paper, we compute the PDOS and phonon dispersions for these compounds along the high-symmetry path (Z-A-M-G-Z-R-X-G) in the first Brillouin zone. The dynamic stability is confirmed by the results, which show the positive phonon modes. Sound velocity is determined by the tilt of the acoustic branch, which also gives phonons a linear dispersion at long wavelengths. Shorter wavelengths are known as optical modes. At the Γ-point, acoustic phonon frequencies reach zero, signifying stability. According to the PDOS, branches of high energy from Ga and Zn predominate for optical properties, but low-frequency acoustic phonons are crucial for heat transmission [46].

## 3.2 Electronic properties

### *3.2.1 Band Structures and Density of States*

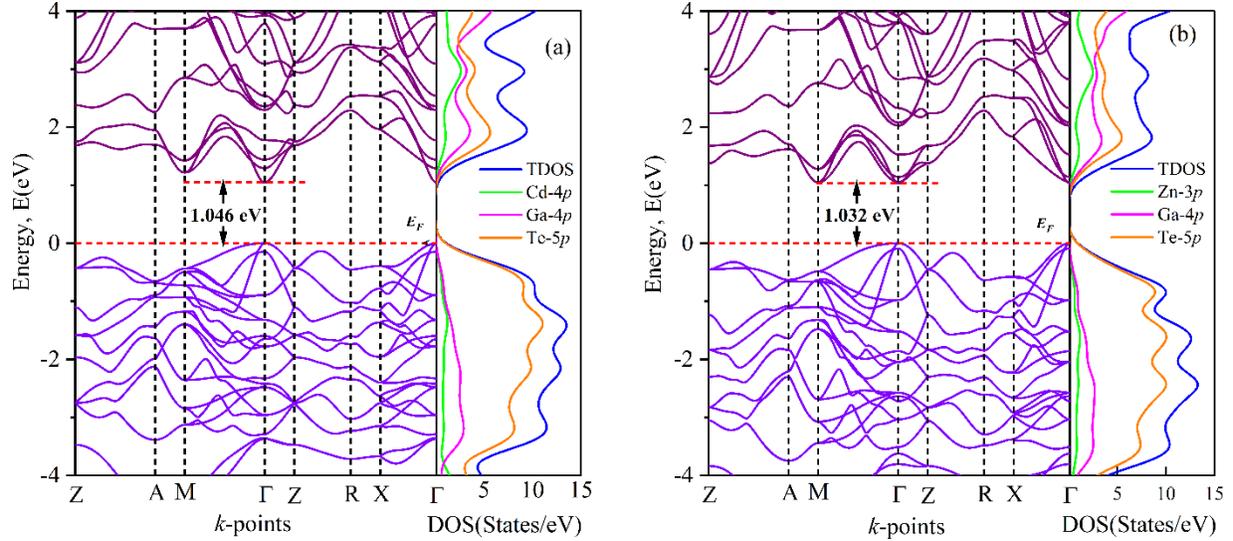

**Figure 3**: The energy band structures of (a) CdGa$_2$Te$_4$ & (b) ZnGa$_2$Te$_4$ calculated by GGA-PBESol approximation.

Electronic band structure is a fundamental quantity in discovering and explaining many physical behaviors such as chemical bonds, electronic transport, superconductivity, optical response, and magnetic order. The Fermi-level bands are critical to the electronic properties and modeling of electrical devices and nanomaterials; therefore, it requires that effective charge-carrier masses be determined accurately and band topologies chosen appropriately. Understanding the electronic band structure in detail is crucial for applications such as catalysts, battery materials, magnetism, or superconducting materials [47][48]. The electronic band gap of such compounds is capable of being adjusted over the visible light region, which is very important for color LEDs. The band structures of CdGa$_2$Te$_4$ and ZnGa$_2$Te$_4$ were investigated at high symmetry positions in a first zone, that is, from -4 to +4 eV. The band structure for CdGa$_2$Te$_4$ and ZnGa$_2$Te$_4$ is presented in **Figures 3(a)** and **3(b)**, respectively, which are calculated on the basis of the GGA-PBESol functional. The electronic band gap is direct, which is observed from the valence band maximum (VBM) and conduction band minimum (CBM) position on the Γ point (0, 0, 0) of the Brillouin zone. Band gaps are 1.046 and 1.032 eV, respectively, for CdGa$_2$Te$_4$ and ZnGa$_2$Te$_4$ with direct bandgaps.

An understanding of these electronic properties becomes useful for the analysis of transport characteristics, such as effective masses of carriers, and is very important to assess solar cell performance of a material. The effective masses of electrons and holes were obtained from the curvature of the electronic band dispersion near the conduction band minimum (CBM) and valence band maximum (VBM) by following **Eq. 4**.

$$\frac{1}{m*} = \frac{1}{\hbar} \frac{\partial^2 E(k)}{\partial K^2} \qquad (4)$$

where E(k) is the energy dispersion in terms of k-space and ℏ is the reduced Planck constant.

**Table 3.** Calculation of the effective mass of electrons ($m_e^*$), holes ($m_h^*$), electron carrier concentration $N_c$ ($cm^{-3}$), hole carrier concentrations $N_v$ ($cm^{-3}$) and exciton binding energy ($E_b$ in meV), Bohr radius ($R_{ex}$ in Å), Exciton temperature ($T_{ex}$ in K).

| Compound | $m_e^*$ | $m_h^*$ | $N_c \times 10^{18}$ | $N_v \times 10^{18}$ | $E_b$ | $R_{ex}$ | $T_{ex}$ |
|---|---|---|---|---|---|---|---|
| CdGa$_2$Te$_4$ | $0.214 m_0$ | $-0.89 m_0$ | 2.24 | 2.14 | 18.85 | 34.3 | 218.0526 |
| ZnGa$_2$Te$_4$ | $0.39 m_0$ | $-0.85 m_0$ | 6.35 | 1.80 | 26.81 | 23.0 | 311.1364 |

Zn has an even greater effective carrier mass than Cd, essentially showing that Cd is better to serve as photovoltaic applications owing to its lower intrinsic effective carrier masses. This is due to the fact that a lower carrier effective mass indicates an increased movement of holes and also higher charge carrier mobility [49]. The masses were derived from a parabolic fit of energy versus wave vector at the band edges [50]. The expected effective mass values of electrons and holes are summarized in **Table 3**, where all the masses are given with respect to the free electron mass (m₀). Heavy (more flat bands) charge carriers with high localization in **Figure 3** are compared with light ones (less flat) that should be very mobile. In order to use the PV devices effectively, materials with a high dielectric constant are required. A low dielectric constant can cause undesirably high charge carrier recombination, which can negatively impact the overall performance of the device [51]. Thus, the characterization of dielectric functions is essential for photovoltaics and optoelectronics.

The DOS is the measure of available electronic states per energy unit and volume unit. The theory predicts, in case of high DOS at some energy level, there are extra places to accommodate. Total density of states (TDOS) and partial density of state (PDOS) were calculated in order to gain an insight into the electronic properties of CdGa$_2$Te$_4$ and ZnGa$_2$Te$_4$ compounds. **Figure 3** displays TDOS and PDOS (GGA-PBESol) results, where the Fermi level is represented with a vertical dashed line at zero energy.

**Table 4:** Comparative DFT band gap values for CdGa$_2$Te$_4$ and ZnGa$_2$Te$_4$ from multiple studies illustrating the consistent Cd > Zn gap trend.

| DFT Study/Year | Material | Functional | Band Gap (eV) | Band Gap Type |
|---|---|---|---|---|
| Present Work (2025) | ZnGa$_2$Te$_4$ | GGA-PBESol | 1.032 | Direct |
| | CdGa$_2$Te$_4$ | GGA-PBESol | 1.046 | Direct |
| | ZnGa$_2$Te$_4$ | HSE06 | 1.895 | Direct |

|  | CdGa$_2$Te$_4$ | HSE06 | 2.21 | Direct |
|---|---|---|---|---|
| Kumar et al. (2017) [52] | ZnGa$_2$Te$_4$ | mBJ | 1.61 | Direct |
|  | CdGa$_2$Te$_4$ | mBJ | 1.78 | Direct |
| Govindaraj et al. (2022) [53] | ZnGa$_2$Te$_4$ | GGA-PBE | 1.01 | Direct |

The lack of a finite value at the Fermi level in the TDOS confirmed the nonmetallic character of these compounds, determining that there were no electronic states available at the Fermi energy, supporting their semiconducting behavior. And the consistency of band gaps from TDOS and band structure clearly identifies them as semiconductors. PDOS analysis indicates that the contributions of Cd-4$p$, Zn-4$p$, Ga-4$p$, and Te-5$p$ orbitals contribute to the TDOS around $E_F$, and the dominant states in the valence band and conduction band are contributed by Te-5$p$ orbitals. The substitution of Cd with Zn has little effect on the TDOS and PDOS because there are few contributions from Zn-4$s$ near the Fermi level, which only causes a small decrease in the band gap of ZnGa$_2$Te$_4$. The hybridized Te, Ga, Zn, and Cd states dominate the overall electronic behavior, which indicates that those compounds have a covalent nature of bonding and structural stability.

## 3.3 Optical characteristics

The optical properties of a material are determined by how its charge carriers interact with incident photons or electromagnetic waves at the material's surface. Recent advancements in science and technology have sparked considerable interest in the optical examination of solids. Optical materials are widely employed in reconfigurable photonics, solar cells, lasers, photodetectors, sensors, and display systems. The optical properties of a material dictate its interaction with electromagnetic radiation. Optoelectronic devices predominantly respond to visible light. These investigations aim to predict materials that will perform effectively in various optoelectronic applications. The energy-dependent optical properties frequently analyzed include the dielectric function ε(ω), refractive index n(ω), optical conductivity σ(ω), reflectivity R(ω), absorption coefficient α(ω), and energy loss function L(ω), where ω = 2 π f denotes the angular frequency of the electromagnetic wave. In this section, we explore how these parameters respond to incident photon energy. The complex dielectric function is expressed as follows in (**Eqs. 5-11**).

$$\varepsilon(\omega) = \epsilon_1(\omega) + i\epsilon_2(\omega) \tag{5}$$

Kramers-Kronig relationships connect the imaginary and real parts. All additional optical constants of importance are given by the following relationships [54].

$$n(\omega) = \sqrt{\frac{|\varepsilon(\omega)| + \epsilon_1(\omega)}{2}} \tag{6}$$

$$k(\omega) = \sqrt{\frac{|\varepsilon(\omega)| - \epsilon_1(\omega)}{2}} \tag{7}$$

$$R(\omega) = \frac{(n-1)^2 + k^2}{(n+1)^2 + k^2} \tag{8}$$

$$\alpha(\omega) = \frac{2k\omega}{c} \tag{9}$$

$$L(\omega) = Im\left(\frac{-1}{\varepsilon(\omega)}\right) = \frac{\epsilon_2(\omega)}{\epsilon_1^2(\omega) + \epsilon_2^2(\omega)} \tag{10}$$

$$\sigma(\omega) = \sigma_1(\omega) + i\sigma_2(\omega) \tag{11}$$

In this section, we specifically examine how CdGa$_2$Te$_4$ and ZnGa$_2$Te$_4$ semiconductors respond to incident electric fields applied along the [100] and [001] directions.

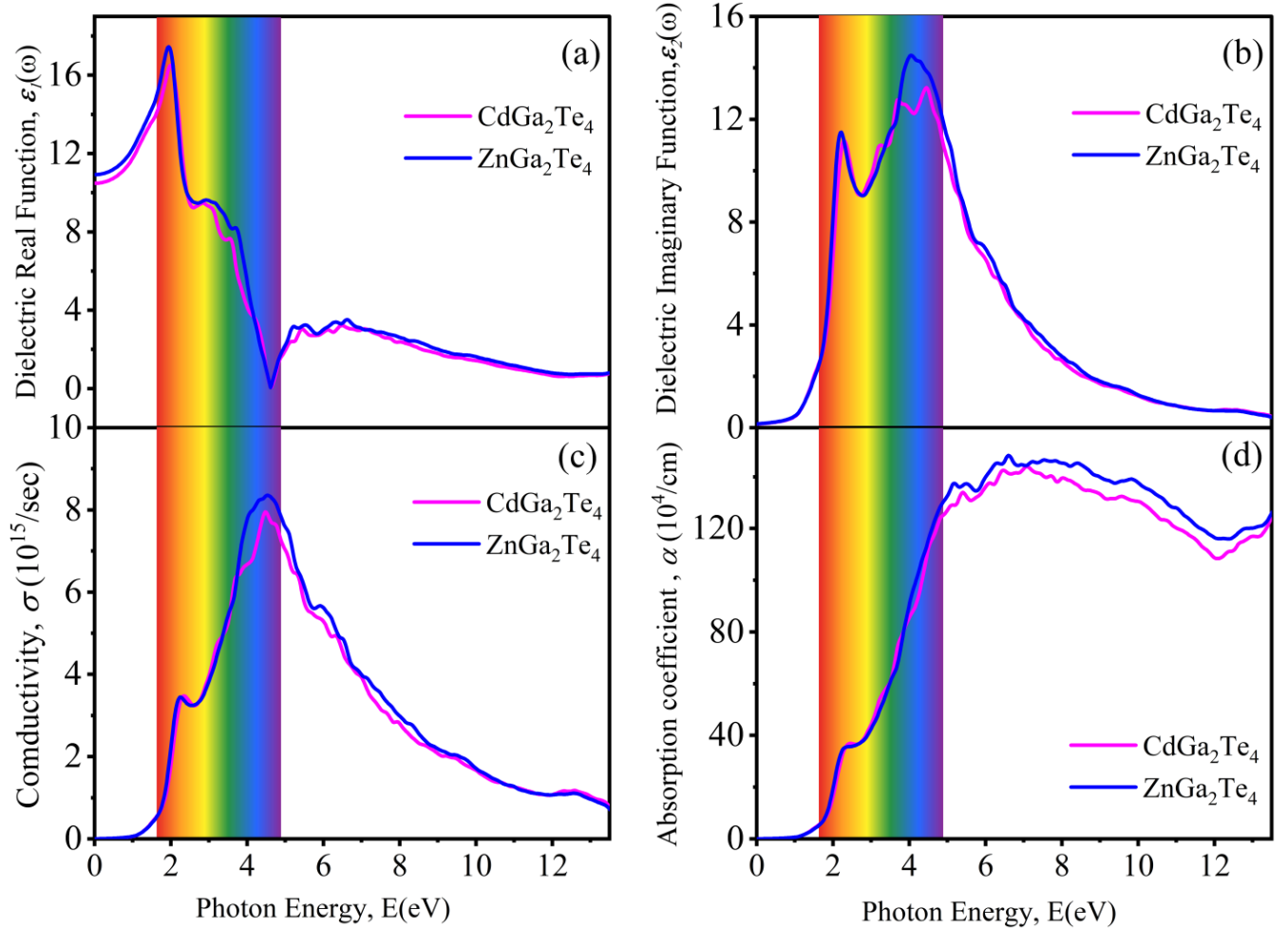

**Figure 4.** (a) Real part of the dielectric function, (b) imaginary part of the dielectric function, (c) conductivity, and (d) absorption coefficient of the CdGa$_2$Te$_4$ and ZnGa$_2$Te$_4$ semiconductors.

The interaction between a material and incident electromagnetic waves is characterized by the dielectric function ε(ω). This function is closely linked to the electronic band structure. **Figures 4(a)** and **4(b)** illustrate the real and imaginary components of the dielectric function for both compounds. The real component is associated with anomalous dispersion and electrical polarization, while the imaginary component pertains to the dissipation of electromagnetic wave energy within the medium. Both compounds exhibit optical isotropy, with negligible variation in the imaginary part. The real component of both compounds begins to decrease gradually after 2.2 eV, whereas the imaginary component starts to decrease after 4.2 eV. The real components suggest the non-metallic nature of the compounds.

Optical conductivity, σ(ω), describes the conduction of free charge carriers over a specific range of photon energies. Due to the presence of a band gap, as indicated by the band structure, photoconductivity commences at approximately 1 eV photon energy, as depicted in **Figure 4c** for both compounds. This observation aligns with the band structure and density of states (DOS). When the incident energy surpasses 1 eV, the materials absorb photons, leading to an increase in photoconductivity and, consequently, electrical conductivity.

The absorption coefficient α(ω) quantifies the extent to which light of a given energy (wavelength) can penetrate a material before being absorbed, providing insights into optimal solar energy conversion efficiency. As shown in **Figure 4d**, absorption begins around 1 eV, consistent with the band structure, and indicates the non-metallic nature of both compounds. Between 6 and 12 eV, these compounds absorb ultraviolet rays most effectively. The optical absorption spectra of CdGa$_2$Te$_4$ and ZnGa$_2$Te$_4$ reveal their strong potential in photovoltaic applications. Both compounds exhibit a steep absorption edge in the visible region, indicating direct optical transitions suitable for efficient solar energy harvesting. The absorption coefficient reaches values on the order of $10^5$ cm$^{-1}$, which is sufficiently high to ensure effective photon capture and reduced thickness requirements for device fabrication. ZnGa$_2$Te$_4$ shows slightly higher absorption than CdGa$_2$Te$_4$ across much of the visible-UV region, suggesting enhanced light-matter interaction and better suitability for thin-film solar cells. Overall, the strong and broad absorption in the visible range confirms that these materials are excellent candidates for optoelectronic and photovoltaic applications.

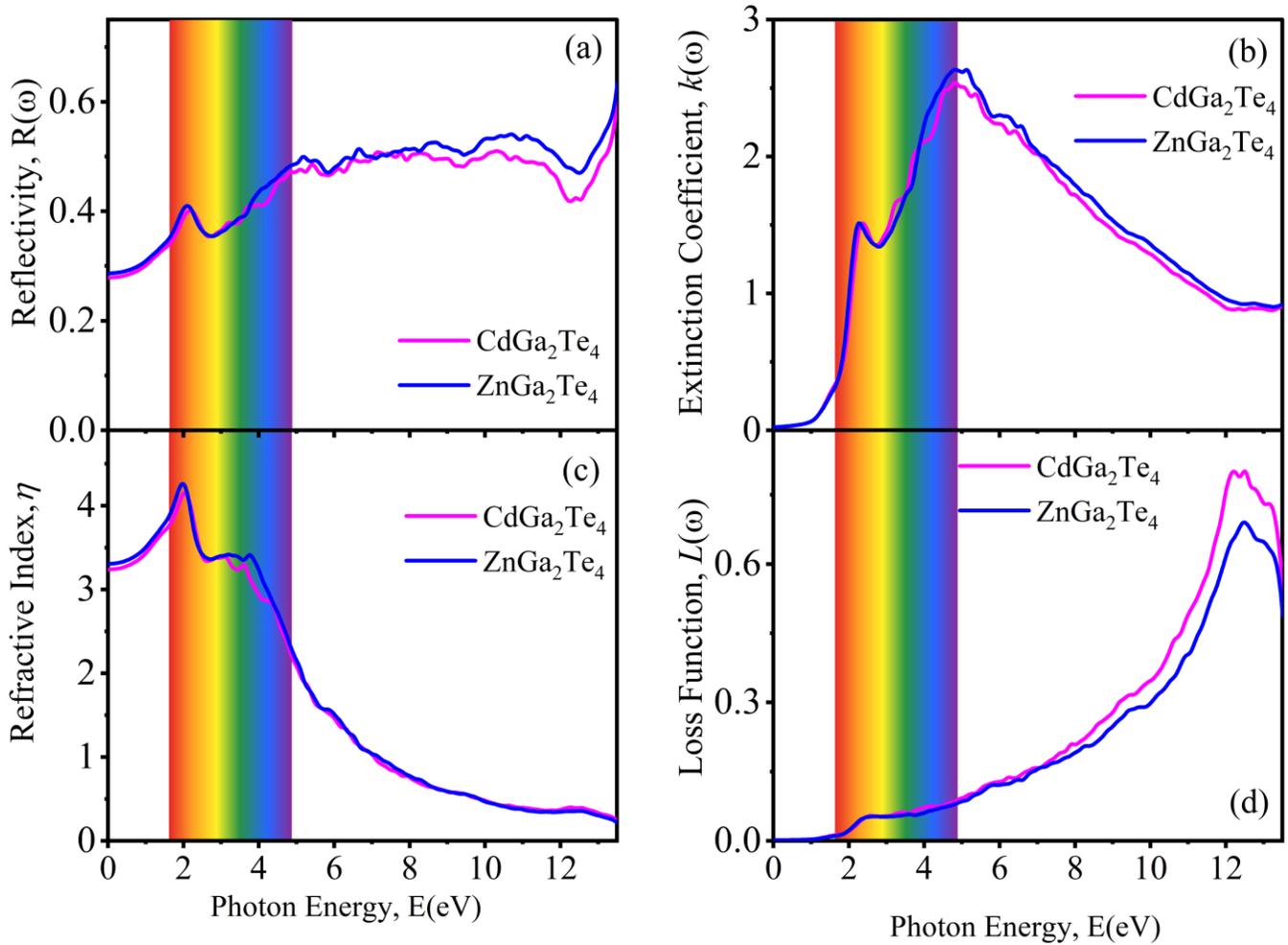

**Figure 5. (a)** Reflectivity, (b) Extinction coefficient, (c) Refractive index, and (d) Loss function of the CdGa$_2$Te$_4$ and ZnGa$_2$Te$_4$ semiconductors.

Reflectivity R(ω) is defined as the ratio of the energy of the wave reflected from the surface to the energy of the wave incident on the surface. **Figure 5a** illustrates the reflectivity of the two compounds. The reflectivity of these compounds was approximately 28% in the infrared (IR) region and increased to approximately 60% at 12–14 eV. These materials are suitable for use as moderate-level reflectors of ultraviolet (UV) rays.

The imaginary component of the refractive index, denoted as η, represents the attenuation of an electromagnetic wave as it traverses a medium, whereas the real component describes the phase velocity of the electromagnetic wave at varying energy levels. In **Figure 5c**. The amount of absorption loss that occurs when an electromagnetic wave, like light, travels through a material is described by the extinction coefficient k(ω) depicted in **Figure 5b**. The extinction coefficient gradually increased from 1 to 4.2 eV, after which it began to decline. The real components of both compounds demonstrated isotropic characteristics and began to decrease beyond 3.9 eV. **Figure 5d** illustrates the loss functions L(ω) for both compounds. Rapid electron passage through a material can excite collective charge oscillation modes, resulting in energy loss by electrons, as evidenced by the peaks in the loss function L(ω) [55,56]. Notable loss peaks were observed at 12 eV, indicating a plasmon energy. At these specific energies, plasma oscillations arise from the collective oscillations of electrons in the jellium model. Significantly, the plasma resonance energies corresponded to marked decreases in the reflectance and absorption coefficients.

## 3.4 Excitonic properties

### 3.4.1 Exciton binding energy

The exciton binding energy ($E_b$) in semiconductors represents the energy needed to dissociate an exciton into its constituent electron and hole. A lower $E_b$ indicates weaker Coulombic attraction between the electron–hole pair, which makes their separation easier. This facilitates more efficient absorption of sunlight and enhances photovoltaic conversion. In this context, the exciton binding energy for $CdGa_2Te_4$ and $ZnGa_2Te_4$ was estimated using the Following **Eq. 12** [57]

$$E_b \approx \frac{13.56 \cdot m_r^*}{m_e \varepsilon_0^2}, \quad m_r^* = \frac{m_e^* m_h^*}{m_e^* + m_h^*} \tag{12}$$

where $m_r^*$ stands for the decreased effective masses of electrons and holes. The electron's rest mass is shown by $m_e^*$, and the static dielectric constant is denoted by $\varepsilon_0$. **Table 3** displays the exciton binding energies for the $ZnGa_2Te_4$ and $CdGa_2Te_4$ compounds. The compounds $CdGa_2Te_4$ and $ZnGa_2Te_4$ have $E_b$ values of 20.06 and 30.51 meV, respectively, which are both within an acceptable range and imply that they could find application in solar systems. Many of the most advanced materials, including $MAPbI_3$ (2–75 meV, 38 meV) [58], closely matched our $E_b$ analysis for $ZnGa_2Te_4$ and $CdGa_2Te_4$ compounds.

The exciton Bohr radius ($R_{ex}$) can be estimated using the Bohr radius of a hydrogen atom ($r_B$).

$$R_{ex} = \frac{m_0}{\mu^*} \varepsilon_0 n^2 r_B \tag{13}$$

$$T_{ex} = \frac{E_b}{K_b} \tag{14}$$

The minimum size of an exciton occurs when the exciton energy level (n) is 1. In general, photovoltaics are more efficient at converting solar power into electricity when electron-hole pairs can more easily dissociate, which happens when the binding energy ($E_b$) is smaller. According to the W-M model, the calculated values of $E_b$ and $R_{ex}$ for $XGa_2Te_2$ compounds (where X = Cd and Zn) are shown in Table 3. Among these, $CdGa_2Te_4$ is found to be a more promising semiconductor compared to $ZnGa_2Te_4$ in terms of exciton performance. Exciton dissociation is facilitated by photon acceleration, and it occurs most effectively when the binding energy is in the range of 25 meV < $E_b$ < 100 meV, with an exciton radius ($R_{ex}$) greater than 15 Å. Furthermore, both $CdGa_2Te_4$ and $ZnGa_2Te_4$ chalcogenide compounds are considered promising materials for solar cell applications due to their low $E_b$ values. The estimated values of $T_{ex}$ are listed in Table 3. Here, the excitonic temperature refers to the maximum temperature at which an exciton may remain stable.

## 3.5 Photovoltaic performance analysis of CdGa$_2$Te$_4$ and ZnGa$_2$Te$_4$ absorbers.

### 3.5.1 SCAPS-1D numerical simulation

The photovoltaic (PV) performance of the heterostructure CdS/XGa$_2$Te$_4$/Cu$_2$O (X = Cd or Zn) has been simulated using SCAPS-1D models to determine device efficiency and performance. CdS served as the electron transport layer (ETL), with a thickness of 100 nm, and Cu$_2$O served as the hole transport layer (HTL), with a thickness of 200 nm. The absorber materials, CdGa$_2$Te$_4$ and ZnGa$_2$Te$_4$, were simulated with Cd$^+$ and Zn$^+$ cations and kept at a constant thickness of 1500 nm. **Figure 6** shows the structural scheme of the solar cell and the corresponding band alignment of the cell. Consistently, the absorber acceptor concentration and defect density were set at $2 \times 10^{16}$ cm$^{-3}$ and $1.772 \times 10^{13}$ cm$^{-3}$, respectively. Simulations were performed under the standard AM 1.5G spectrum at an illumination of 1000 mw cm$^{-2}$ and at an operating temperature of 300 K under typical in situ exposure of the solar cell materials in real applications. The major material parameters of CdS, XGa$_2$Te$_4$ (Cd and Zn), and Cu$_2$O are tabulated in **Table 5**, the respective energy band alignments of relevance in efficient charge transport and recombination at device interfaces. To determine the attributes related to the quality of a solar cell, the continuity equation for both electrons and holes is utilized. Additionally, the program includes Shockley-Read-Hall (SRH) recombination statistics to simulate the device's operation following **Eqs. 15-17.**

$$\frac{d^2}{dx^2}\psi(x) = \frac{q}{\epsilon_0 \epsilon_r}[p(x) - n(x) + N_D - N_A + \rho_p - \rho_n] \qquad (15)$$

Here, $\psi$ signifies the electric potential, $q$ stands for the electronic charge, $\epsilon_0$ indicates the permittivity in a vacuum, and $\epsilon_r$ denotes the relative permittivity. Furthermore, N$_A$ and N$_D$ denote the density of acceptors and donors, respectively. The variables p and n represent the concentrations of holes and electrons, respectively. Finally, $\rho_p$ and $\rho_n$ denote the charge density of the hole and the electron, respectively.

$$\frac{\delta p}{\delta t} = \frac{1}{q}\frac{\delta J_p}{\delta x} + (G_p - R_p) \qquad (16)$$

$$\frac{\delta n}{\delta t} = \frac{1}{q}\frac{\delta J_n}{\delta x} + (G_n - R_n) \qquad (17)$$

The equations shown feature the variables $J_p$ and $J_n$ denoting the current density of holes and electrons, respectively. Furthermore, $G_n$ and $G_p$ signify the generation rates of electrons and holes, respectively, whereas $R_n$ and $R_p$ refer to the electrons and recombination rate of holes, respectively. The SCAPS 1D software can generate a steady-state response in a single dimension, considering the given equations and boundary conditions.

**Table 5:** Fundamental parameters of each layer of the device.

| Parameters | n- (CdS) [59] | p-(ZnGa$_2$Te$_4$) | p-(CdGa$_2$Te$_4$) | p$^+$- (Cu$_2$O)[60] |
|---|---|---|---|---|
| Layer | ETL | Absorber | Absorber | HTL |
| Thickness (nm) | 100 | 1500 | 1500 | 200 |
| Bandgap, $E_g$ [eV] | 2.400 | 1.03 | 1.047 | 2.200 |
| Electron affinity, $\chi$ [eV] | 4.200 | 4.5 | 4.34 | 3.400 |
| Dielectric permittivity (relative) | 10.000 | 10.90 | 10.8 | 7.500 |
| Effective DOS at CB [cm$^{-3}$] | 2.2×10$^{18}$ | 6.35×10$^{18}$ | 2.24×10$^{18}$ | 1×10$^{19}$ |
| Effective DOS at VB [cm$^{-3}$] | 1.8×10$^{19}$ | 1.80×10$^{19}$ | 2.14×10$^{19}$ | 1×10$^{19}$ |
| Electron thermal velocity (cms$^{-1}$) | 10$^7$ | 1×10$^7$ | 1×10$^7$ | 1×10$^7$ |
| Hole thermal velocity (cms$^{-1}$) | 10$^7$ | 1×10$^7$ | 1×10$^7$ | 1×10$^7$ |
| Electron Mobility, $\mu_n$ [cm$^2$V$^{-1}$s$^{-1}$] | 100 | 2250 | 800 | 2×10$^2$ |
| Hole mobility, $\mu_p$ [cm$^2$V$^{-1}$s$^{-1}$] | 25 | 650 | 200 | 8.6×10$^3$ |
| Donor density, $N_D$ [cm$^{-3}$] | 10$^{17}$ | 10 | 10 | 0 |
| Acceptor density, $N_A$ [cm$^{-3}$] | 10 | 2×10$^{16}$ | 2×10$^{16}$ | 1×10$^{18}$ |
|  | Neutral | Single Donor | Single Donor | Single Donor |
| Total defect density, $N_t$ [cm$^{-3}$] | 1.772×10$^{17}$ | 1.772×10$^{13}$ | 1.772×10$^{13}$ | 1×10$^{14}$ |

The values in **Table 5** represent the parameters of the HTL, ETL, and absorbers that were used to simulate the modeled solar cell. Doping density, defect density, and thickness are important variables to optimize the performance metric of the cell.

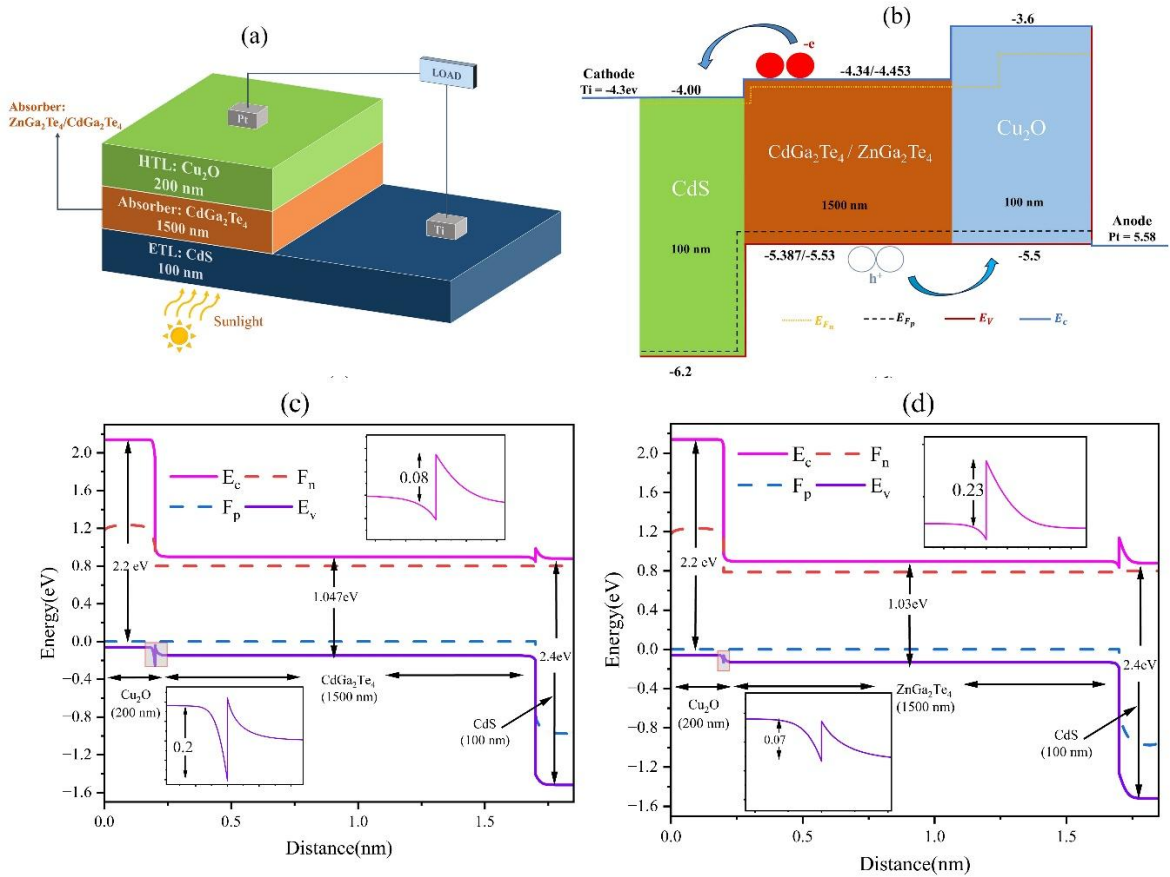

**Figure 6:** (a) Cell structure of the CdGa₂Te₄ and ZnGa₂Te₄ absorber, (b) device structure, (c) band diagram for the CdGa₂Te₄ absorber, and (d) band diagram for the ZnGa₂Te₄ absorber of an optimized solar cell structure.

### 3.5.2 *Energy band alignment and band offsets*

The band alignment of the energy gaps between the electron transport layer (ETL), the absorber material, and the hole transport material (HTL) is a crucial factor in maximizing the overall efficiency of solar cells. Band alignment not only dictates charge extraction and transport efficiency but also plays an important role in suppressing recombination losses, both of which directly impact device performance [61,62]. To achieve high-performance SCs, it is necessary to maintain a very small conduction band offset (CBO) for efficient electron injection, along with a sufficiently wide valence band offset (VBO) to effectively inhibit hole recombination [63,64]. In the present work, the band alignments of the ETL CdS, combined with $XGa_2Te_4$ (X = Cd, Zn) as absorber layers and $Cu_2O$ as the HTL, were studied systematically to determine the most favorable material combination for optimized charge separation and improved power conversion efficiency. The following equations were used for calculating the conduction band offset (CBO) and valence band offset (VBO) by following **Eqs. 18-19**.

$$CBO = \chi_{ETL} - \chi_{abs} \quad (18)$$

$$VBO = E_{g,ETL} - E_{g,abs} - CBO \quad (19)$$

Where $\chi_{abs}$ and $\chi_{ETL}$ represent the electron affinities of the perovskite materials and ETL materials, respectively, and $E_{g,abs}$ and $E_{g,ETL}$ are their bandgaps.

**Figure 6(b)** shows the band alignment of both Pt/CdS/CdGa₂Te₄/Cu₂O/Ti and Pt/CdS/ZnGa₂Te₄/Cu₂O/Ti structures. Here, in the CdS/CdGa₂Te₄/Cu₂O structure, CdS has a conduction band minimum (CBM) of –4.00 eV and a valence band maximum (VBM) of –6.2 eV. The CBM and VBM of the absorber are –4.3 eV and –5.38 eV, respectively. In this case, the conduction band offset (CBO) was found to be 0.14 eV, while the valence band offset (VBO) was much higher, at about 1.34 eV. A higher VBO than CBO means that the valence band edge of the absorber is significantly lower than that of the buffer layer, creating a barrier for hole transport [65]. On the other hand, the CBM and VBM of the HTL material are –3.6 eV and –5.5 eV, respectively, which are lower than those of the ETL. Although CdS is a well-established ETL material, the relatively larger CBO reduces the efficiency of electron injection, introducing increased resistive losses in the cell.

In the CdS/ZnGa₂Te₄/Cu₂O structure, the CBM and VBM of the ETL and HTL remain unchanged, while the absorber shows notable differences. The ZnGa₂Te₄ absorber has a CBM and VBM of –4.453 eV and –5.53 eV, respectively, which are slightly higher than those of CdGa₂Te₄. Using the given formulas, the calculated CBO and VBO are about 0.3 eV and 1.07 eV, respectively. An absorber with a slightly higher positive CBO typically shows superior electron injection and reduced recombination losses compared to one with a lower or negative CBO, thus balancing recombination blocking with efficient carrier transport [66]. Therefore, the structure with a ZnGa₂Te₄ absorber is expected to exhibit better electron injection and reduced recombination losses than the CdGa₂Te₄-based structure. The MBO energy is significant for maintaining the smoother gradients of the energy levels and thus improving the carrier transport path. Schottky energy barriers of 0.08 eV and 0.2 eV, depicted in Figure 6(c) at the CdS/CdGa₂Te₄ and CdGa$_2$Te$_4$/Cu$_2$O interfaces, and of 0.23 eV and 0.07 eV shown in Figure 6(d) for the CdS/ZnGa$_2$Te$_4$ and ZnGa$_2$Te$_4$/Cu$_2$O interfaces, effectively prevent holes from enhancing recombination with the transport layer. Such barriers are intended to suppress the interface recombination and, in principle, can improve the performance of the device.

### 3.5.3 *J-V characteristics and photovoltaic performance*

The current density voltage (J–V) response is a fundamental tool for assessing PSC performance because it directly reflects device parameters such as short-circuit current density (J$_{sc}$), open-circuit voltage (V$_{oc}$), fill factor (FF), and power conversion efficiency (PCE) [67]. Typically, J–V curves display a diode-like signature, where forward bias causes a sharp current rise, while reverse bias significantly suppresses current flow, indicating efficient carrier collection and limited recombination [68]. These measurements remain the primary benchmark for solar cell optimization. The diode-like J–V behavior of these cells can be modeled using the Shockley diode formula in the following **Eq. 20** [69].

$$J = J_0(\exp\left(\frac{qV}{nk_BT}\right) - 1) \qquad (20)$$

where $J_0$ is the reverse saturation current, $q$ is the electron charge, $V$ is the applied voltage, $n$ is the ideality factor, $k_B$ is Boltzmann's constant, and $T$ is the temperature. This model helps analyze device efficiency based on the product of current, voltage, and fill factor, normalized by the input power. The power conversion efficiency (PCE) is determined using the following **Eq. 21** [70].

$$PCE = \frac{J_{SC} \cdot V_{OC} \cdot FF}{P_{in}} \qquad (21)$$

where $P_{in}$ denotes the incident light power density. This expression evaluates the overall efficiency of a solar cell by relating the product of the short-circuit current, open-circuit voltage, and fill factor to the total input power.

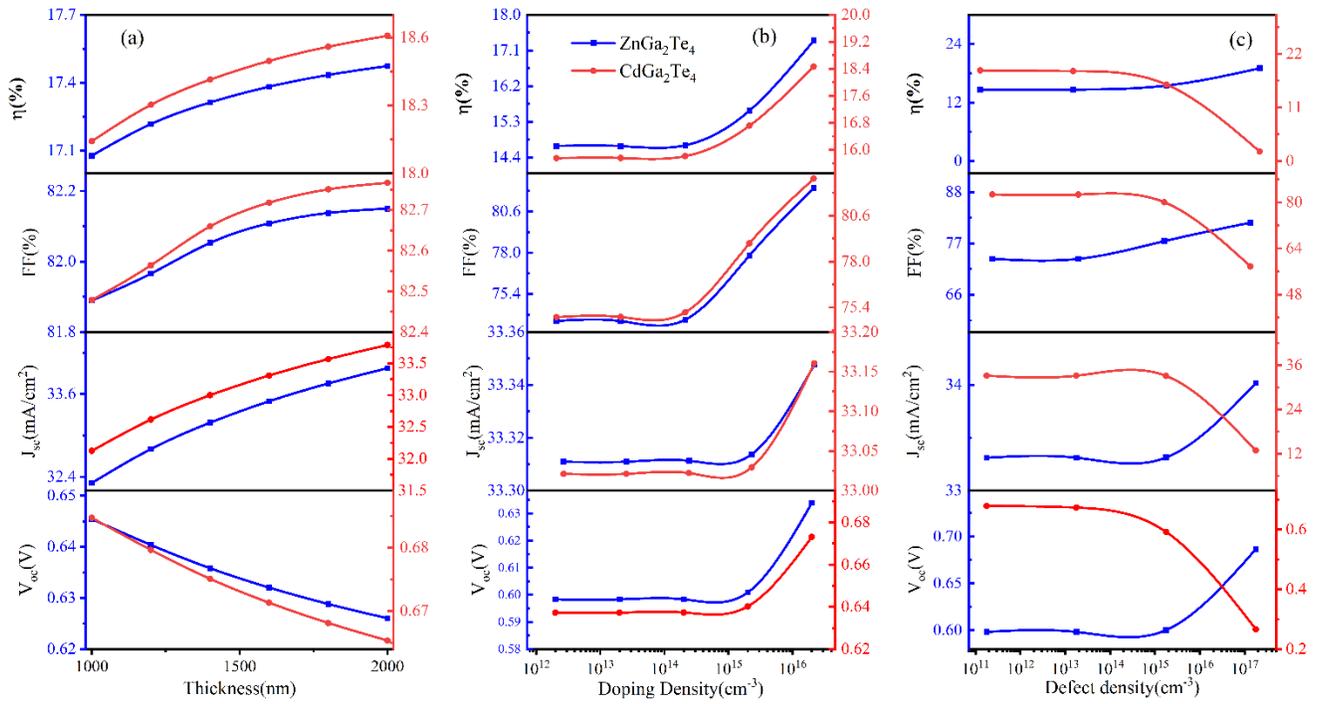

**Figure 7**: Effect of varying (a) thickness, (b) doping density, and (c) defect density of both absorbers on the solar cell performance.

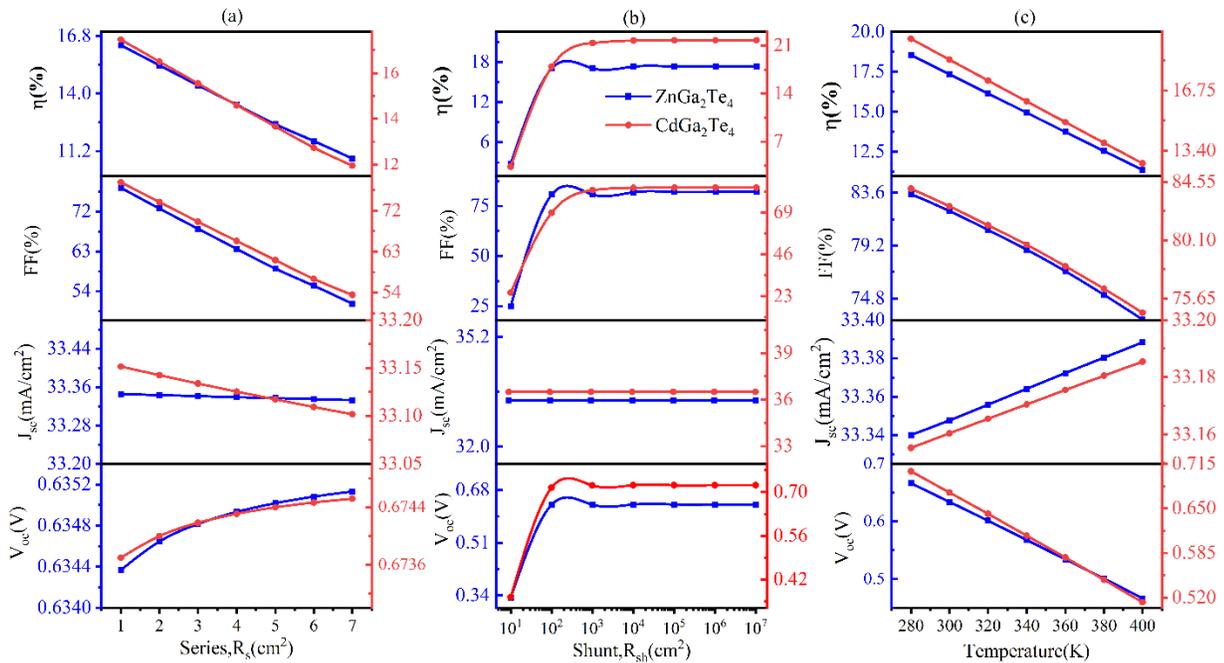

**Figure 8**: Effect of varying (a) series resistance $R_s$, (b) shunt resistance Rsh, and (c) temperature of both absorbers on the solar cell performance.

The performance curve in **Figure 7(a) reveals** that the efficiency (η), short-circuit current density (Jsc), and fill factor (FF) all improved with increasing thickness values. In each case, the Cd-based absorber maintained better outcomes. Although the open-circuit voltage (Voc) dropped with increasing thickness for both absorbers, $CdGa_2Te_4$ still maintained slightly higher values than $ZnGa_2Te_4$. When the doping density increased, both materials showed big improvement in each performance metric depicted in **Figure 7(b)**. Above $10^{14}$ cm$^{-3}$, η and FF started to increase rapidly, while $V_{OC}$ started its rapid increase right after crossing $10^{15}$ cm$^{-3}$. Uniquely, the JSC curve dropped for both absorbers for a short range of doping density and started to increase rapidly beyond that short range. On the other hand, the effect of varying defect density on solar performance was exhibited in **Figure 7(c)**. $ZnGa_2Te_4$ stayed more stable even when a lot of defects were introduced. In contrast to $CdGa_2Te_4$, the Zn-based compound showed slow (η and FF) as well as rapid ($V_{OC}$ and $J_{SC}$) improvement in its curves. Meanwhile, the $CdGa_2Te_4$ compound's η, FF, JSC, and $V_{OC}$ started to drop quickly. The effect of series resistance shown in **Figure 8(a)** caused η, FF, and Jsc to decrease steadily for both absorbers, whereas both $V_{OC}$ increased with respect to the increment of resistance. Shunt resistance had a considerable effect on all performance metrics, which is very well depicted in **Figure 8(b)**. But after reaching a certain value, all 4 performance parameters leveled off. Under the applied conditions, $ZnGa_2Te_4$ had a more stable Jsc and Voc than $CdGa_2Te_4$. Both curves of JSC remained unchanged throughout the process. Finally, **Figure 8(c)** shows that raising the working temperature caused η, FF, and Voc to drop significantly for both devices. However, Jsc showed a linear and steady increase with rising temperature. The Cd-based structure couldn't dictate with its superior performance in each parametric variation. However, the optimized device simulated results suggest that the overall performance of the $CdGa_2Te_4$ absorber is slightly better than that of $ZnGa_2Te_4$.

**Table 7**: Photovoltaic result comparison with previous literature works.

| Structure | $V_{oc}$ (V) | Jsc (mA/cm²) | FF (%) | Eta (%) | Ref. |
|---|---|---|---|---|---|
| Pt/Cu$_2$O/CdGa$_2$Te$_4$/CdS/Ti | 0.6731 | 33.160474 | 82.69 | 18.46 | Present Work |
| Pt/Cu$_2$O/ZnGa$_2$Te$_4$/CdS/Ti | 0.6339 | 33.347635 | 82.08 | 17.35 | Present Work |
| Al/FTO/CdS/CIGS/BaSi$_2$/Mo | 0.843 | 40.56 | 76.76 | 26.24 | [71] |
| Ag/Cu$_2$O/K$_2$TlSbI$_6$/TiO$_2$/FTO | 0.81 | 45.07 | 82.21 | 31.77 | [72] |
| Ag/Cu$_2$O/K$_2$TlAsI$_6$/TiO$_2$/FTO | 0.93 | 40.69 | 84.07 | 30.01 | [72] |
| Cu/ZnO:Al/i-ZnO/n-CdS/p-CMTS/Pt | 0.883 | 34.41 | 83.74 | 25.43 | [73] |
| Cu/ZnO:Al/i-ZnO/n-CdS/p-CMTS/p+-SnS/Pt | 1.074 | 36.21 | 81.04 | 31.51 | [73] |
| FTO/In$_2$S$_3$/CdTe | 0.6312 | 25.505526 | 82.37 | 13.26 | [74] |
| FTO/In$_2$S$_3$/CdTe/FeSi$_2$ | 0.656 | 49.776 | 83.68 | 27.35 | [74] |

The comparison table clearly shows this work achieves competitive efficiency. Although the efficiencies of this work are slightly lower, it is still significant because they consist of a relatively simple structure, and the analyzed materials are not expensive or rare earth. The fill factors were found to be very high and favorable for practical usage.

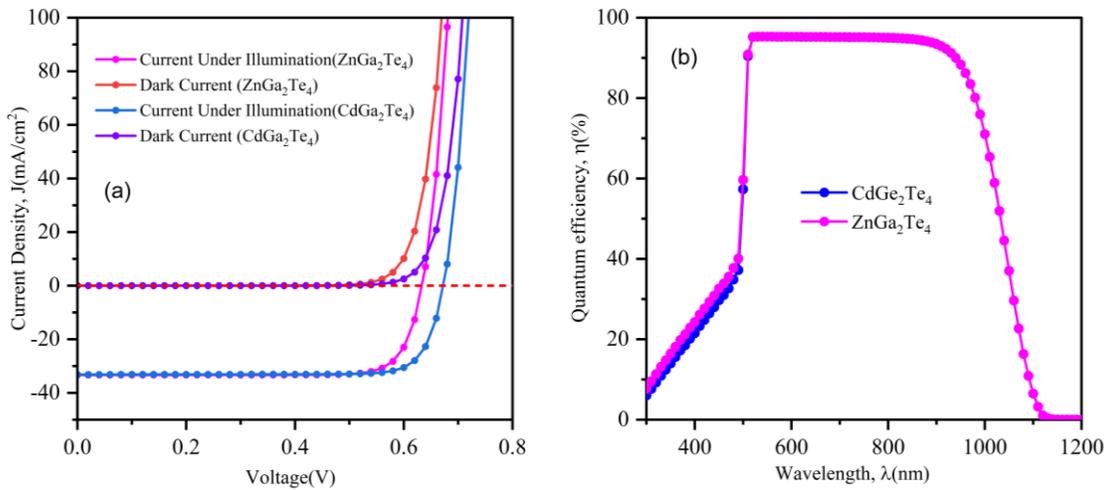

**Figure 9**: (a) J-V characteristics and (b) quantum efficiency of both absorbers (CdGa$_2$Te$_4$ and ZnGa$_2$Te$_4$).

### 3.5.4 Quantum Efficiency and I-V characteristic

The I–V characteristics were simulated under both dark and illuminated conditions, depicted in **Figure 9(a)**. The results indicate that both compounds exhibit a pronounced increase in current density with increasing voltage. Under illumination, the Zn-based compound demonstrated a higher current density than the Cd-based compound, suggesting enhanced photocurrent generation. Although the current density J (mA/cm²) for $CdGa_2Te_4$ also increased, it did not rise as sharply as that of $ZnGa_2Te_4$. Under dark conditions, both compounds showed an exponential increase in current density, characteristic of forward bias. The more pronounced increase in dark current for $ZnGa_2Te_4$ implies a higher reverse saturation current, whereas the lower dark current in $CdGa_2Te_4$ suggests reduced recombination loss, potentially leading to improved solar cell performance. Quantum efficiency (QE) is a vital metric for evaluating the capability of a photovoltaic device to convert incident light into charge carriers. It represents the fraction of incoming photons that successfully generate carriers, providing critical insight into absorber performance across the solar spectrum [80]. **Figure 9(b)** illustrates the external QE spectra of $CdGa_2Te_4$- and $ZnGa_2Te_4$-based solar cells employing $Cu_2O$ as the HTL and CdS as the ETL. Measurements were conducted over the 300–1200 nm range, covering both visible and near-infrared (NIR) wavelengths. The device configurations show a high QE for both compounds. Both devices maintained a QE exceeding 90% within the wavelength range of 510–930 nm. Beyond the 930 nm wavelength, the QE declined significantly. The Cd-based device achieved a maximum QE of 95.3% at approximately 530 nm. Similarly, the Zn-based device also exhibited a peak of 95.3% at 530 nm. This observation affirms the close similarity of the QE spectra for $ZnGa_2Te_4$ and $CdGa_2Te_4$, as both compounds display nearly identical QE values throughout the visible region, with only minor variations arising from slight bandgap differences.

### 3.5.5 Final optimization result

For the $CdS/ZnGa_2Te_4/Cu_2O$ cell, with a $J_{sc}$ of 33.3 mA/cm², a $V_oc$ of 0.6339 V, and an FF of 82.08%, we obtained a PCE of 17.35%. $CdGa_2Te_4$ devices demonstrated higher performances. The Jsc, Voc, and FF of 33.160474 mA/cm², 0.6731 V, and 82.69%, respectively, were subsequently obtained from the $CdS/CdGa_2Te_4/Cu_2O$ cell types, leading to a PCE of 18.46%. It was found that the maximum output power could be obtained when the thickness of $ZnGa_2Te_4$ and $CdGa_2Te_4$ both reach 1500 nm. The thicknesses of the ETL (CdS) and HTL ($Cu_2O$) were maintained at 100 nm and 200 nm, respectively, for all materials. The $CdGa_2Te_4$ absorber was more promising in terms of the fill factor and efficiency. Furthermore, the Voc of $CdGa_2Te_4$ was superior to that of $ZnGa_2Te_4$. The Cd-based system had lower Jsc in comparison to the Zn-based system, with the only exception of the Jsc. After simulations, the doping levels of $ZnGa_2Te_4$ and CdGa$_2$Te$_4$ are set as 2×10$^{16}$, and both defect densities are set as 1.772×10$^{13}$ to achieve optimized results.

## 4. Conclusions

In the present work, we investigated the electronic, optical, and excitonic properties of CdGa$_2$Te$_4$ and ZnGa$_2$Te$_4$ using density functional theory (DFT). Analysis of their electronic structure revealed direct band gaps at 1.046 eV and 1.032 eV, respectively, where the density of states showed a strong involvement of Te-5p and Ga-4p orbitals in the semiconducting properties of such compounds. Estimations of effective mass and carrier concentration indicate good charge transport properties towards electronic devices. The optical response, absorption, and conductivity in particular are in good agreement with their electronic structure, underlining the potential of these materials for optoelectronic applications. For the excitonic parameters of low binding energy, large Bohr radius, and stable excitonic temperature, we observe efficient exciton dissociation in these compounds, which therefore qualify these materials as candidates for solar cell applications. Device-level simulations of Pt/Cu$_2$O/CdGa$_2$Te$_4$/CdS/Ti and Pt/Cu$_2$O/ZnGa$_2$Te$_4$/CdS/Ti heterostructures have also shown high-quality photovoltaic behavior with fill factors (82.69% and 82.08%) compared to current densities of about 33 mA/cm². The CdGa$_2$Te$_4$-based device showed the best performance in terms of open-circuit voltage (0.6731 V) and efficiency (18.46%) compared to ZnGa$_2$Te$_4$-based devices (open-circuit voltage of 0.6339 V and efficiency of 17.35%). Overall, CdGa$_2$Te$_4$ and ZnGa$_2$Te$_4$ are promising efficient absorbers of next-generation PV and optoelectronic applications.

**Conflict of interest**

The authors declare that they have no known competing financial interests or personal relationships that could have appeared to influence the work reported in this article.

**Acknowledgement**

The authors acknowledge the support from the computational facilities at University of Rajshahi.

**CRediT Author Statement**

**Md Hasan Shahriar Rifat:** Conceptualization, Methodology, Software, Data curation, Supervision, Validation, Writing - Original Draft, Writing - Review & Editing. **Tanvir Khan:** Formal analysis, Investigation, Writing -

Original Draft. **Md Arafat Hossain Shourov:** Formal analysis, Investigation, Writing - Original Draft. **Md Sahat Bin Sayed:** Formal analysis, Investigation, Writing - Original Draft. **Md. Saiful Islam:** Validation, Investigation, Writing - Original Draft, Writing - Review & Editing.

**Data Availability Statement**

The data that support the findings of this study are not publicly available because no suitable public repository is currently available to host these files. Reasonable requests for access to the data will be considered by the corresponding author.